\documentclass[12pt]{iopart}
\usepackage{iopams}
\usepackage{graphicx}
\usepackage{amssymb}
\usepackage{epstopdf}
\usepackage{bm}

\usepackage{color}

\newcommand{\new}{\textcolor{magenta}}
\begin{document}

\title{Energy and momentum transfer in one-dimensional trapped gases by stimulated light scattering}

\author{
N. Fabbri$^{1,2}$ , C. Fort$^{1,2}$, M. Modugno$^{3,4}$, S. Rosi$^{1,2}$, and M. Inguscio$^{1,5}$
}
\address{
$^1$ LENS European Laboratory for Non-linear Spectroscopy, and Dipartimento di Fisica e Astronomia - Universit\`a di Firenze, I-50019 Sesto Fiorentino, Italy\\
$^2$ CNR-INO Istituto Nazionale di Ottica, I-50019 Sesto Fiorentino, Italy\\
$^3$ Dpto. de F\'isica Te\'orica e Historia de la Ciencia, Universidad del Pa\'is Vasco UPV/EHU, 48080 Bilbao, Spain\\
$^4$ IKERBASQUE, Basque Foundation for Science, 48011 Bilbao, Spain\\
$^5$ INRIM Istituto Nazionale di Ricerca Metrologica, I-10135 Torino, Italy
}

\ead{fabbri@lens.unifi.it}

\begin{abstract}
In ultracold atoms settings, inelastic light scattering is a preeminent technique to reveal static and dynamic properties at nonzero momentum. In this work, we investigate an array of one-dimensional trapped Bose gases, by measuring both the energy and the momentum imparted to the system via light scattering experiments. The measurements are performed in the weak perturbation regime, where these two quantities -- the energy and momentum transferred -- are expected to be related to the dynamical structure factor of the system. We discuss this relation, with special attention to the role of in-trap dynamics on the transferred momentum.
\end{abstract}

\submitto{\NJP}
\maketitle
\tableofcontents

\section{Introduction}
Stimulated scattering of light or particles from condensed-matter systems -- solids, liquids, and gases -- is a powerful tool for providing fundamental insight into the structure of matter. Elastic scattering of x-ray photons has permitted to disclose the atomic order and electron distribution in crystalline solids, as well as the arrangement of atoms in molecules \cite{wollanRMP1932}.
Similarly, inelastic neutron scattering has unveiled the phonon spectrum of superconductors and the superfluidity of liquid helium \cite{bramwellNATMAT2014}.

In cold atomic systems, inelastic scattering of photons -- also known as Bragg spectroscopy --  has been used to study harmonically trapped three-dimensional (3D) Bose-Einstein condensates (BECs) \cite{stengerPRL1999, kozumaPRL1999, steinhauerPRL2002,richardPRL2003}, strongly interacting BECs across a Feshbach resonance \cite{pappPRL2008}, and strongly interacting fermions \cite{veeravalliPRL2008,paganoNP2014}, through direct observation of the net momentum imparted to the system. The transferred momentum  is easily measured in this kind of settings, since the atomic density distribution, observed after time-of-flight in the far-field regime, directly reflects the in-trap momentum distribution. Instead, cold atoms in optical lattices have been investigated by measuring the increase of energy following a modulation of the lattice amplitude where the excitation has zero momentum \cite{stoferlePRL2004}, and with scattering experiments where the excitation has non-zero momentum \cite{clementPRL2009,fabbriPRL2012,fabbriarXiv2014}. The energy of a condensate, even in the presence of shallow lattices, is easily extracted from the time-of-flight density distribution of the gas \cite{dalfovoRMP1999}, whereas the energy of strongly-interacting systems realized in deep optical lattices -- as a Mott insulator -- is not directly accessible, unless with single-site resolution experiments \cite{shersonNAT2010}. In the case of deep optical lattices, the energy excess produced by the Bragg perturbation can be measured by lowering the lattice depth, {\it i.e.}, driving the system in a less interacting regime, and let it thermalize \cite{stoferlePRL2004,clementPRL2009}.

In the linear response regime, both energy and momentum transfer are related to the dynamic structure factor \cite{brunelloPRA2001}, which carries key information about the dynamical behaviour and correlations of the system.
However in trapped condensates, while the energy is a conserved quantity, momentum is not conserved due to the presence of the trap. Thus, if the Bragg pulse duration is not negligible compared to the inverse of the trap frequency, the momentum imparted by the Bragg beams can be affected by the in-trap dynamics \cite{brunelloPRA2001,blakiePRA2002}, complicating the connection between the momentum transferred and the dynamical structure factor. On the other hand, a short Bragg pulse would result in a limited spectral resolution.

In this work, we use inelastic light scattering for accessing the dynamical structure factor of an array of one-dimensional (1D) Bose gases. The dimensionality of the system plays a crucial role: in 1D quantum systems, correlations -- which directly reflect on the dynamical structure factor -- bring to peculiar phenomena, such as fermionization of strongly-interacting bosons \cite{paredesNAT2004,kinoshitaSCI2004,hallerSCI2009}, or spin-charge separation of interacting fermions \cite{giamarchibook}, which do not have any higher-dimensional equivalent. Moreover, for 1D systems, the mechanisms and characteristic times of thermalization are currently under investigation \cite{kinoshitaNAT2006,langenNP2013}. This may affect the measurement of the energy transferred via light scattering. On the other side, momentum measurement of 1D trapped gases may be influenced by the in-trap dynamics, as above mentioned.
The purpose of this work is to investigate in experiment the relation between energy and momentum imparted to an array of 1D gases due to Bragg scattering, in a typical regime of parameters \cite{clementPRL2009,fabbriarXiv2014}, and to discuss the effect of the in-trap dynamics on the transferred momentum.

The paper is organized as follows. In Sec. \ref{Sec:DEDP0} we focus on the comparison of the response of the array of 1D gases to the scattering experiments in terms of energy deposited and momentum boost imparted to the system. We present the experimental setup and discuss the results, obtained in a regime of weak perturbation that is well described in the framework of the linear response theory. We also directly compare the susceptibility of this system to that one of a 3D non-interacting condensate. In Sec. \ref{Sec:dyn} we study the effect of the in-trap dynamics on the momentum transfer by recording the evolution of the response of the system in time, after the Bragg excitation.

\section{Energy and momentum transferred to an array of 1D gases}
\label{Sec:DEDP0}

\subsection{Experimental setup}
We produce an array of 1D gases by loading a Bose-Einstein condensate of $^{87}$Rb in a two-dimensional optical lattice created by two mutually orthogonal laser standing waves of wavelengths $\lambda_L = 765$ nm. The loading is performed with an exponential ramp of $t_r=$250 ms, with time constant $t_r/3$. The final depth of the lattice is $V_L = 30 E_r$, with $E_r= h^2/(2m\lambda_L^2)$, $m$ being the atomic mass and $\lambda_L$ the lattice wavelength. This value is chosen to be high enough to freeze the transverse degree of freedom of each 1D gases (the radial trapping frequency is $\omega_\bot=2\pi\times(42\pm2)$ kHz), and suppress the tunneling of particles between different tubes on the timescale of the experiment.

The equilibrium state of the system is completely described by two dimensionless parameters \cite{kheruntsyanPRA2005}, that is, {\it (i)} the interaction parameter $\gamma=m g_{1D}/(h^2 \rho)$, where $g_{1D}$ is the one-dimensional interaction strength \cite{olshani}, and {\it (ii)} the reduced temperature $t=2m k_B T/(\hbar \rho)^2$, which depend on density $\rho$ and temperature $T$.

The significant parameters that characterize the array can be estimated by rescaling the interparticle interaction strength $g_{3D}$ as in Ref. \cite{fabbriarXiv2014}. We estimate the array to consist of about $4\times 10^3$ 1D tubes, and the central tube to have  $\gamma\simeq 1$ and $t\simeq1$, with density $\rho\simeq 5$ $\mu$m$^{-1}$ and chemical potential $\mu/h\simeq 3.6$ kHz.

The study of this system is carried out by imparting a perturbation to the array of 1D tubes given by two simultaneous off-resonance laser pulses with time duration $t_B=3$ ms, which determines an interaction-time broadening of $\simeq 150$ Hz. Note that $t_B\simeq T/4$, with  $T=2\pi/\omega_x$ being the trap period along the axis of the tubes. The laser light is detuned by 200 GHz from the $^{87}$Rb resonance. The two beams have tunable relative detuning $\Delta\omega$ (up to tens of kHz), and produce a moving Bragg grating with amplitude $V_B=h \times 900$ Hz. The wavevector of the Bragg grating is adjusted to be along the axial direction of the tubes, and it is fixed at $ q=(7.3\pm 0.2)$ $\mu$m$^{-1}$. In the experiments, we vary the energy $\hbar \omega$ of the excitation by tuning $\Delta\omega$ (being $\omega=|\Delta\omega|$) and we measure the response of the system, in terms of energy and momentum transfer.

\subsection{Results}
\label{sec:DEDP}

\begin{figure}[b!]
\flushright
\includegraphics[width= 1\columnwidth]{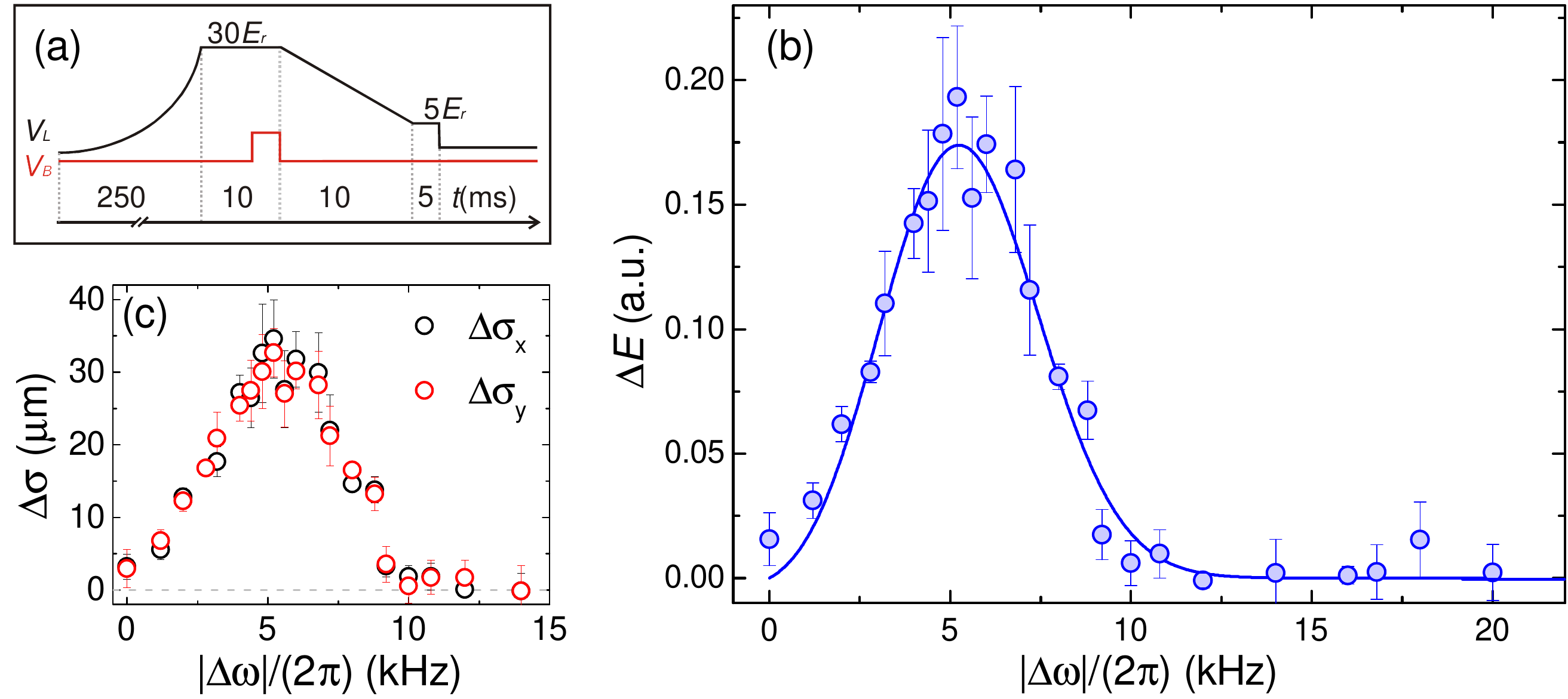}
\caption{{\bf Energy transfer.} (a) Experimental timing used for measuring the energy spectrum. The optical lattice is raised to $30 E_r$ in 250 ms. During the holding time the Bragg beams are shined onto the atoms, then the whole system is let thermalize by lowering the lattice depth to $5E_r$. After $5$ms the trap is released, and the energy is measured from the time-of-flight density distribution. (b) Energy transferred via Bragg scattering to an array of 1D gases in an optical lattice of height $V_L=30$ $E_r$, as a function of the excitation frequency. The signal is obtained from the squared width of the central peak of the time-of-flight density distribution ($\sigma_x^2 + 2 \sigma_y^2$). The fitting curve is given by a gaussian function multiplied by the frequency $\omega G(\omega)$.  (c) Increase of the {\it rms} size along the axis of the tubes ($\sigma_x$) and along a perpendicular axis ($\sigma_y$), as a function of the excitation frequency.
}\label{DE}
\end{figure}

For measuring the energy transfer, after the Bragg pulse we lower the lattice height to $V_L=5 E_r$, where the whole gas is expected to thermalize in a superfluid regime. After 5ms the gas is released and let expand ballistically for a time-of-flight $t_{tof}= 25$ms, then we record the density distribution of the atomic cloud. The experimental timing is sketched in the Fig. \ref{DE}(a).
From the time-of-flight images, we extract the squared width of the central peak of the resulting interferogram $\sigma^2=\sigma_x^2 + 2 \sigma_y^2$ \cite{nota1} and subtract the background $\sigma_0^2$, corresponding to the value measured in  the absence of the Bragg pulse, in order to obtain the experimental signal.
This quantity is proportional to the energy imparted to the system, as previously demonstrated \cite{fabbriarXiv2014}. The latter, in turns, is related to the dynamical structure factor $S(q,\omega)$ of the system through the relation \cite{zambelliPRA2000}
\begin{equation}
\Delta E (q,\omega) = \left(\frac{2 \pi}{\hbar }\right)\left(\frac{V_{B}}{2}\right)^2 t_B \,\omega S(q,\omega)\,,
\label{Eq:DE}
\end{equation}
valid in the linear response regime.
The measured energy spectrum, normalized to its integral, is shown in Fig. \ref{DE}(b).
In order to verify that the system has thermalized after the Bragg pulse, in Fig. \ref{DE}(c) we plot separately the increase of the {\it rms} size observed in each direction ($\Delta\sigma_x$ and $\Delta\sigma_y$): in spite of the symmetry breaking induced by the Bragg perturbation imparted along the axis of the 1D gases ($x$ direction), $\Delta\sigma_x$ and $\Delta\sigma_y$ show the same dependence on the Bragg frequency, indicating that an efficient thermalization process has occurred during the ramping down of the lattices.

\begin{figure}[ht!]
\begin{center}
\includegraphics[width= 1\columnwidth]{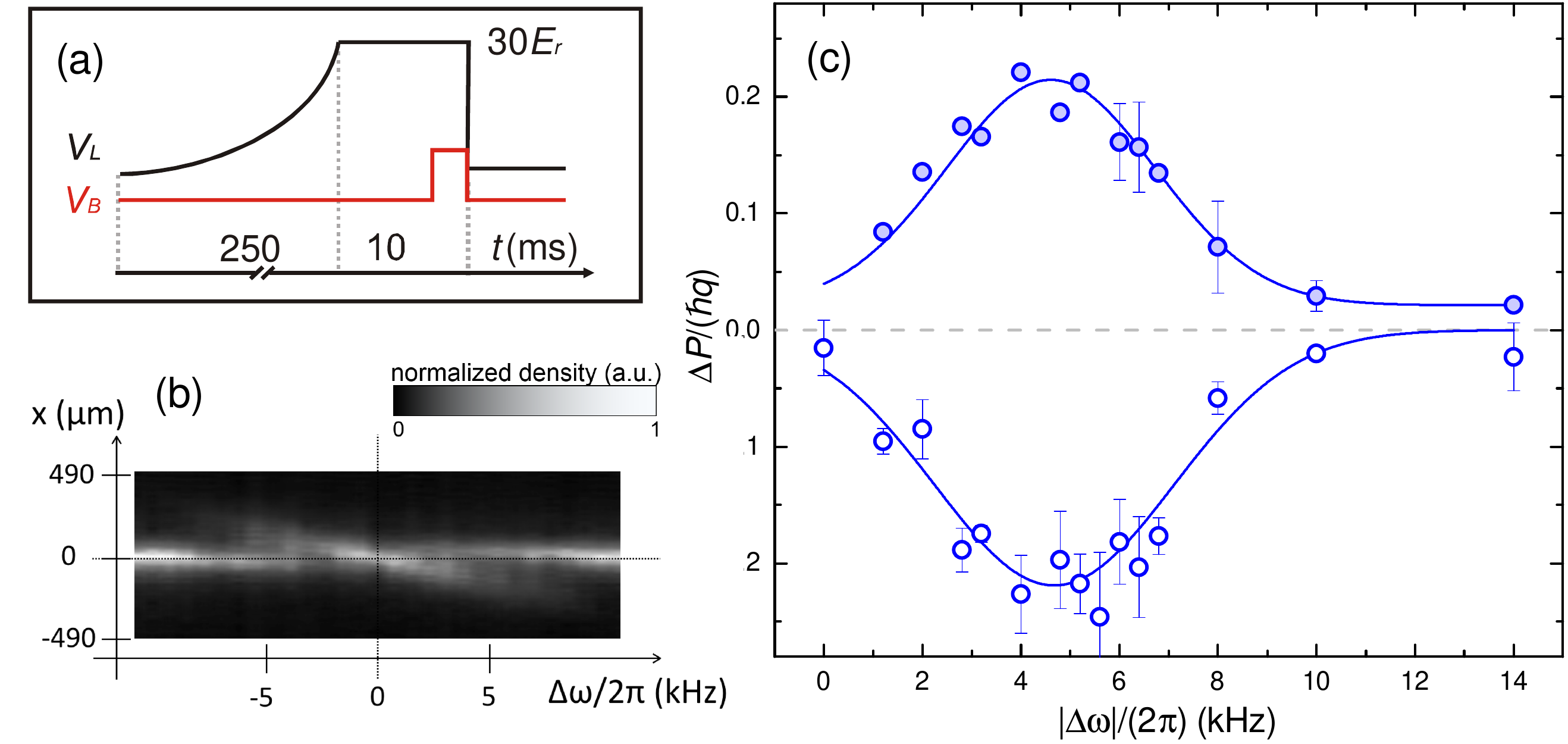}
\end{center}
\caption{{\bf Momentum transfer.} (a) Experimental timing. The optical lattices are abruptly switched off immediately after the Bragg pulse, and the momentum transfer is measured from the center-of-mass shift of the whole cloud. (b) Normalized density profile $n(x)$ along the axis of the 1D gases --  integrated along the line of sight $z$ and the $y$ direction -- for different values of the relative detuning of the Bragg beams $\Delta\omega/(2\pi)$ (false colours). On resonance, the central peak is depleted and the atoms are ejected along the $x$ direction.
(c) Momentum transferred to an array of 1D gases. The filled (empty) dots, at positive (negative) momentum along the axis of the tubes, correspond to $\Delta\omega<0$ ($\Delta\omega>0$). The Bragg parameters of this measurement are the same as described for the measurement of the energy transfer.
}\label{DP}
\end{figure}

In the experiment, we also measure the total momentum imparted to the same system. To this purpose, after the Bragg pulse, the atoms are abruptly released directly from the trap, as represented in Fig. \ref{DP}(a),  so that in-trap momentum is mapped into the atomic density distribution after time-of-flight. When the Bragg perturbation is on resonance, momentum is transferred efficiently, and the time-of-flight images of the density distribution exhibit a small cloud of excited atoms ejected from the main cloud. Figure \ref{DP}(b) shows the evolution of the normalized density profiles $n(x)$ integrated along the line of sight and along the $y$ direction (orthogonal to the axis along which momentum is transferred) with the relative detuning between the Bragg beams.
From the time-of-flight images, the net moment boost $\Delta P(\omega, q)$ is obtained by measuring the displacement of the center of mass -- relative to the unperturbed position -- as \cite{katzPRL2004}
\begin{equation}
\Delta P(q, \omega)=\frac{m}{t_{tof}} \int x\ (n(x,y)-n_0(x,y))\ dx  dy\,,
\label{Eq:DPimage}
\end{equation}
where $n(x,y)$ and $n_0(x,y)$ are the density profiles integrated along the line of sight and normalized to the unity, with and without the Bragg excitation, respectively. The experimental spectra normalized to the momentum of the excitation $\hbar q$, $\Delta P(q, \omega)/(\hbar q)$, are shown in Fig. \ref{DP}(c). Filled (empty) dots correspond to positive (negative) momentum, along the axis of the tubes.

As remarked in Refs. \cite{brunelloPRA2001,blakiePRA2002,tozzoNJP2003}, momentum is a conserved quantity only in the absence of any external trapping potential. In the latter case, for a perturbation in the linear response regime and with $\omega t_B\gg1$, $\Delta P(q,\omega)$ is related to the dynamical structure factor through the following relation
\begin{equation}
\Delta P (q,\omega) = \left(\frac{2 \pi}{\hbar }\right)\left(\frac{V_B}{2}\right)^2 t_B \,q  S(q,\omega).
\label{Eq:DP}
\end{equation}
For a trapped gas with axial trapping frequency $\omega_x$ significantly smaller than the radial one, this equation still holds in a wide range of parameters, provided that $\omega t_B\gg1$ and $\omega_x t_B\ll1$ \cite{tozzoNJP2003}. In the present case, the first condition is well satisfied as $\omega t_B \simeq 80$ on resonance, while we have $\omega_x t_B \simeq 1$, which does not satisfy the second condition. Thus the comparison between the quantities extracted from the measurements of energy and momentum transfer is not straightforward. In order to address quantitatively this issue, we fit $\Delta P(q, \omega)$ with
a gaussian function $G(\omega)$, where  center, width and  amplitude are free parameters, and $\Delta E(q, \omega)$ with $\omega\ G(\omega)$, as follows from Eq. \ref{DE}. The gaussian centers obtained from the measured spectra are respectively $(4.5\pm0.2)$ kHz for the momentum transfer, and $(4.3\pm0.3)$ kHz for the energy transfer, with their widths being $(2.5\pm0.2)$ kHz and $(2.3\pm0.2)$ kHz, respectively.
These results are consistent within the error bars, allowing us to conclude that, with this choice of parameters, both these experimental approaches measure the same quantity.

\subsection{Comparison with the response of a 3D condensate}

\begin{figure}[t!]
\flushright
\includegraphics[width= 1\columnwidth]{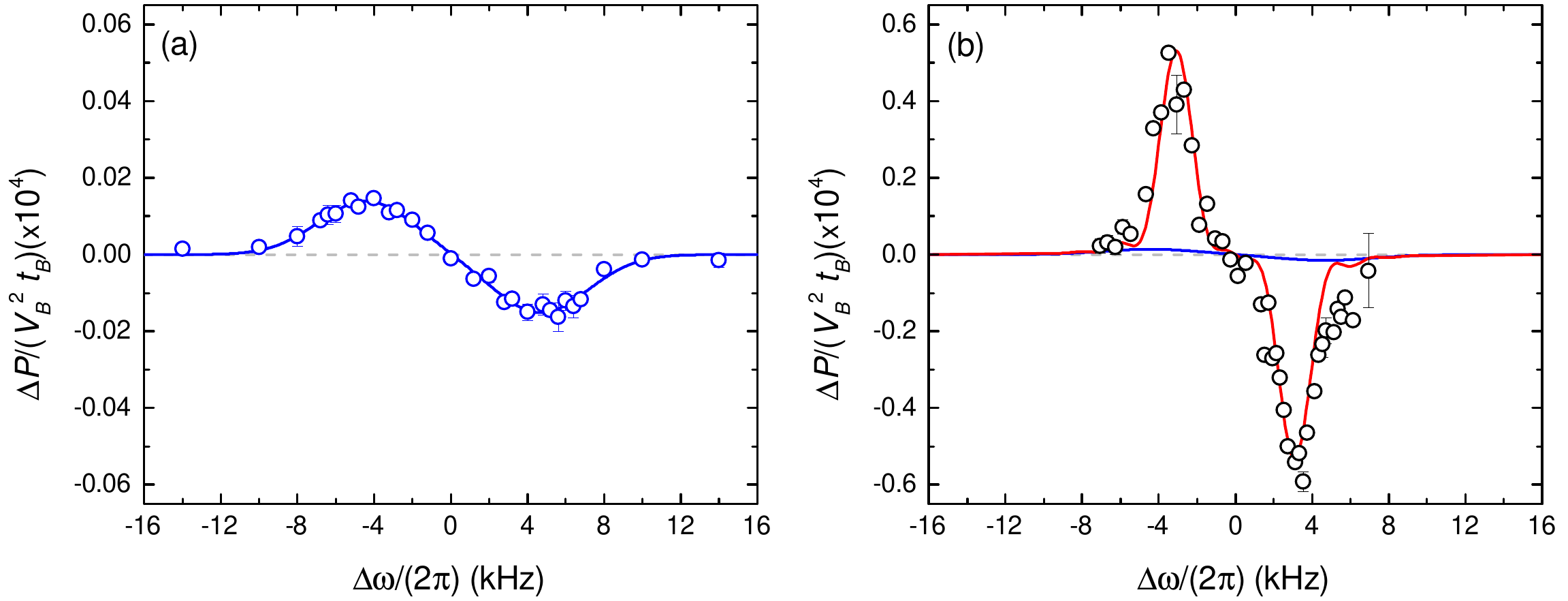}
\caption{{\bf Comparison between the transferred momentum $\Delta P$ to {\bf (a)} an array of trapped 1D gases and {\bf (b)} a non-interacting 3D condensate.} The horizontal scale represents the frequency difference between the Bragg beams. The spectrum of the array of 1D gases has been  obtained using a Bragg parameters $t_B=3$ ms and $V_B= h \times 900$ Hz, whereas the spectrum of the 3D condensate has been obtained with $t_B=0.5$ ms, and $V_B= h \times 540$ Hz. The amplitude of both the spectra has been rescaled by the pulse strength $V_B^2 t_B$ to directly compare the susceptibility of the two systems.  Note that the vertical scale in the first graph is 10 times smaller than in the second one.  A fit of the experimental spectrum of the 1D gases with a sum of two gaussian functions (blue curve) is shown in Fig. \ref{1D3D} (a) and  reported also in Fig. \ref{1D3D} (b) for highlighting the comparison between the response of the two systems. The red continuous line in Fig. \ref{1D3D} (b) is the solution of the time-dependent Schr\"odinger equation for a non-interacting gas, given the value of $V_B/h$, with no free parameters (see text).}
\label{1D3D}
\end{figure}

As a reference system, we also measured the transferred momentum of a 3D expanded condensate,  since its response is well described by a non-interacting model. The interparticle interactions are suppressed indeed by letting the BEC free fall for $5$ ms of time-of-flight before performing the scattering experiment. For direct comparison, in Fig. \ref{1D3D} we show the experimental spectrum of the array of 1D gases obtained as previously described (Fig. \ref{1D3D}(a)), and the spectrum of the 3D non interacting condensate  (Fig. \ref{1D3D}(b)). In both Figs. \ref{1D3D}(a) and 3(b), the signal is normalized to the Bragg strength $V_B^2 t_B$.
In this Figure, we also report the exact solution for a free-particle system (continuous red line), which shows an excellent agreement with the experimental data. This prediction is obtained by solving the Schr\"odinger equation \cite{notaFFT} in the presence of the Bragg potential $V(x,t)=\theta(t-t_B)V_{B}\cos(qx-\omega t)$ (for $t\ge0$), and does not contain any fitting parameter.
From the comparison of the two spectra, we can notice that the response of the 1D tubes is much broader, as a consequence of interactions \cite{fabbriarXiv2014}.
Moreover, the comparison between the two amplitudes indicates that the susceptibility of the 1D tubes is about 35 times lower than the one of the 3D non-interacting condensate.

The very low susceptibility of the array of 1D tubes, with the Bragg parameters that we have used, is a first indication that the response of the system lies in the linear regime of weak perturbation, where the relations in Eq. (\ref{Eq:DE}) and Eq. (\ref{Eq:DP}) are expected to hold. We have also verified the behaviour of the experimental signal as a function of Bragg power in a range that includes the used value, and demonstrate the dependence of the signal on $V_B$ to be quadratic, as expected in the framework of the linear response theory \cite{brunelloPRA2001}.

\section{Effect of in-trap dynamics on the momentum transfer}
\label{Sec:dyn}

As previously discussed, momentum is not a conserved quantity in the presence of a trapping potential. Therefore, the measurement of the momentum following a Bragg excitation can be affected by the in-trap dynamics before the release \cite{blakiePRA2002,tozzoNJP2003}.
For the cases discussed so far, the momentum transferred has been measured immediately after the Bragg pulse, see Fig. \ref{DP}. For the 1D gases, even if $t_B\simeq T/4$, as seen in Sec.\ref{sec:DEDP}, $\Delta P$ and $\Delta E$ carry the same information, then we can infer that no appreciable effect of the dynamics during the Bragg pulse has been observed.

\begin{figure}[t!]
\flushright
\includegraphics[width= 0.8\columnwidth]{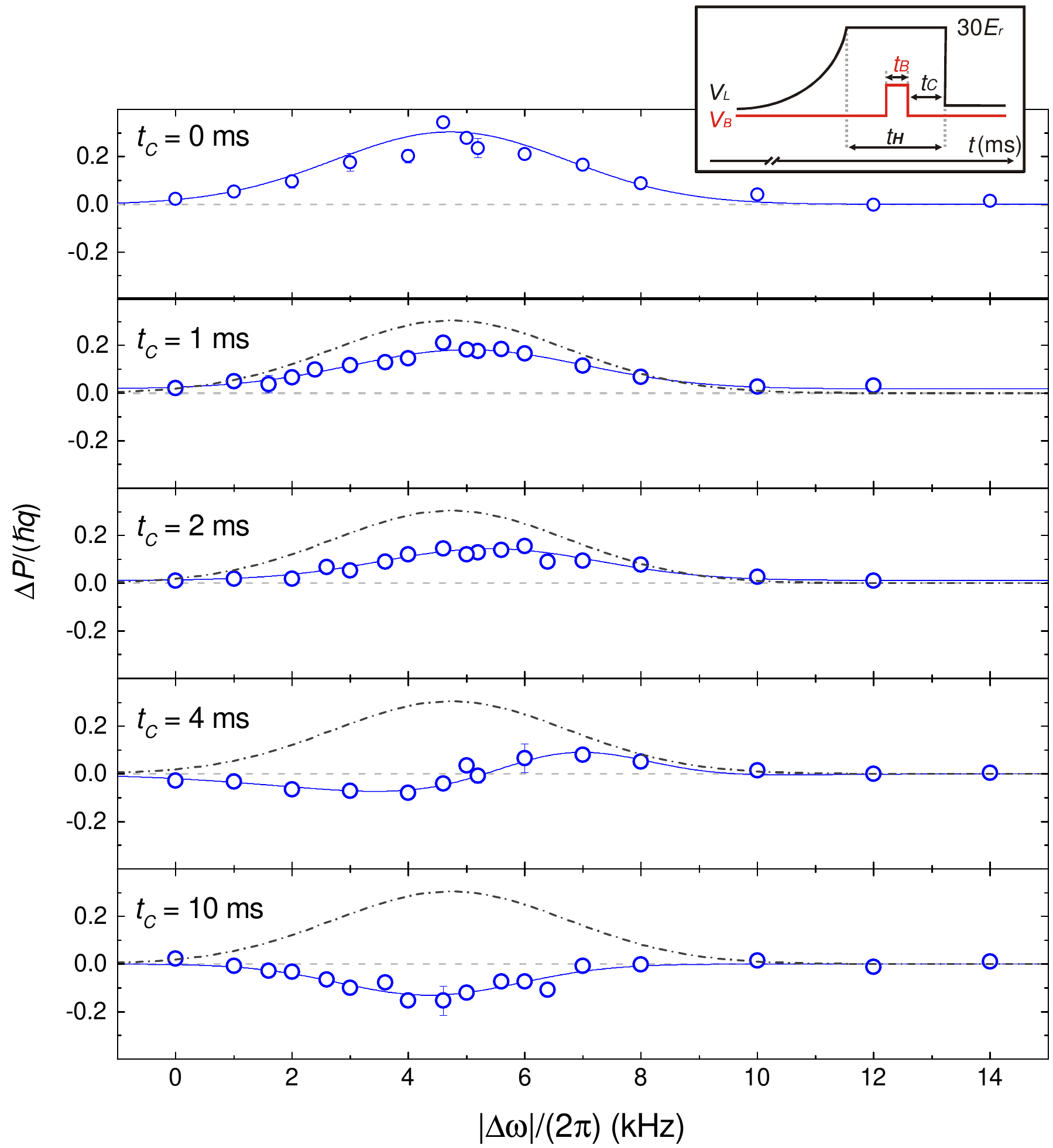}
\caption{{\bf In-trap dynamics after the Bragg excitation.} $\Delta P/(\hbar q)$ is reported as a function of $|\Delta \omega|/(2\pi)$, for different holding times in the trap after the Bragg excitation: $t_C=0,1,2,4,10$ ms. The continuous blue lines are guides for the eyes. The curve corresponding to $t=0$ ms is also reported in the other panels (as a dashed-dotted line) for comparing the signal shapes and amplitudes. Inset: experimental time sequence. The total holding time in the lattices is kept constant ($t_H=30$ ms).
}\label{spettridiversitempi}
\end{figure}

Now, we characterize more in depth the effect of the in-trap dynamics on the response of the 1D gases. To this purpose, we  measure $\Delta P(q, \omega)/(\hbar q)$ at variable time after the end of the Bragg perturbation. In Fig. \ref{spettridiversitempi}(a) we can observe a modification of the system response with time. The Bragg time-duration is fixed to the value $t_B=3$ ms, and the total holding-time of the atoms in the lattice trap ($t_H$) is kept constant, while we vary the time $t_C$ between the end of the Bragg pulse and the release of the trap from 0 ms up to $10$ ms, as sketched in the inset of the figure. During the first $2$ ms after the Bragg pulse, the total spectral weight of the signal undergoes a suppression with time (note that the vertical scale is the same in all the panels of the Figure), which eventually results in a negative amplitude at $t_C=10$ ms.
Remarkably, the shape of the signal at $t_C=4$ ms is asymmetric and qualitatively different from the other cases.

The latter behavior can be understood by considering the effect of the in-trap dynamics {\it during} the Bragg pulse. Let us consider a system of $N$ interacting particles trapped in a harmonic potential, described by the following Hamiltonian
\begin{equation}H =
\sum_{i=1}^{N}\left[\frac{\hat{\bm{p}}_i^2}{2m} + V_{ho}(\bm{x}_i)
+\sum_{j<i} V^{int}(\bm{x}_i-\bm{x}_j)\right].
\end{equation}
The evolution of the total momentum and position operators along each spatial directions can be easily obtained from the Heisenberg equations as $\dot{\hat{p}}_{\alpha}=(-i/\hbar)[\hat{H},\hat{p}_{\alpha}]=m\omega^{2}\hat{x}_{\alpha}$ and $\dot{\hat{x}}_{\alpha}=(-i/\hbar)[\hat{H},\hat{x}_{\alpha}]=-\hat{p}_{\alpha}/m$, with $\hat{O}_{\alpha}=\sum_{i=1}^{N}\hat{O}_{i\alpha}$ ($\alpha=1,2,3$, $\hat{\bm{O}}=\hat{\bm{x}},\hat{\bm{p}}$). For the first relation we have used the fact that $\partial_{x_{j\alpha}}V^{int}_{ij}=-\partial_{x_{i\alpha}}V^{int}_{ij}$. Then, restricting the discussion to the 1D case, it is straightforward to get that the average momentum evolves in the trap as
\begin{equation}
\langle p\rangle(t)= - m \omega_{x} \langle \widehat{x}\rangle_{0} \sin (\omega_x t) + m\langle\dot{\widehat{x}}\rangle_{0} \cos(\omega_{x}t)
\label{ho}
\end{equation}
where $\langle\widehat{x}\rangle_{0}$ and $\langle\dot{\widehat{x}}\rangle_{0}$ are the average values of the density and velocity distributions at time $t=0^+$ immediately after the end of the Bragg pulse.
We remark that this result is valid in general for any interacting system, regardless of the temperature, the statistics (being the particles bosons or fermions) and the dimensionality of the system. In fact, it is a well known result that the dynamics of the center of mass in the presence of harmonic trapping is decoupled from the internal degrees of freedom of the system (see e.g. \cite{dalfovoRMP1999}).

Let us now turn to the effect of the Bragg pulse, that we assume of the form $g(t) V_B \cos(q x-\omega t)$.
First, let us consider an instantaneous Bragg pulse described by $g(t)= \delta(t)$.
After the pulse, at $t=0^{+}$, the density distribution is basically unperturbed ($\langle \widehat{x}\rangle_{0}=0$). Then, the Bragg perturbation only affects the initial velocity distribution $n(\dot{\widehat{x}}_{0})$, so that its mean value is $\langle\dot{\widehat{x}}\rangle_{0}= (N_B(\omega)/N) \hbar q/m$, where $N_B/N$ is the ratio of the number of diffracted atoms to the total number of atoms, and depends on the excitation frequency $\omega/(2\pi)$. As follows from Eq. (\ref{ho}), in this case $\langle p\rangle(t)$ vanishes exactly for $t=T/4$, $T$ being the period of the trap, for any excitation frequency.

Instead, for a finite duration of the Bragg pulse (and in particular, if $t_B$ is comparable with the trap period), even the spatial distribution of the atomic ensemble may undergo modifications during the Bragg perturbation, depending on the excitation frequency.
This makes the initial value of the center-of-mass $\langle\widehat{x}\rangle_{0}$ in Eq. (\ref{ho}) non vanishing and $\omega$-dependent,
therefore affecting the following dynamics and changing the shape of the signal.

\begin{figure}
\flushright
\includegraphics[width= 0.49\columnwidth]{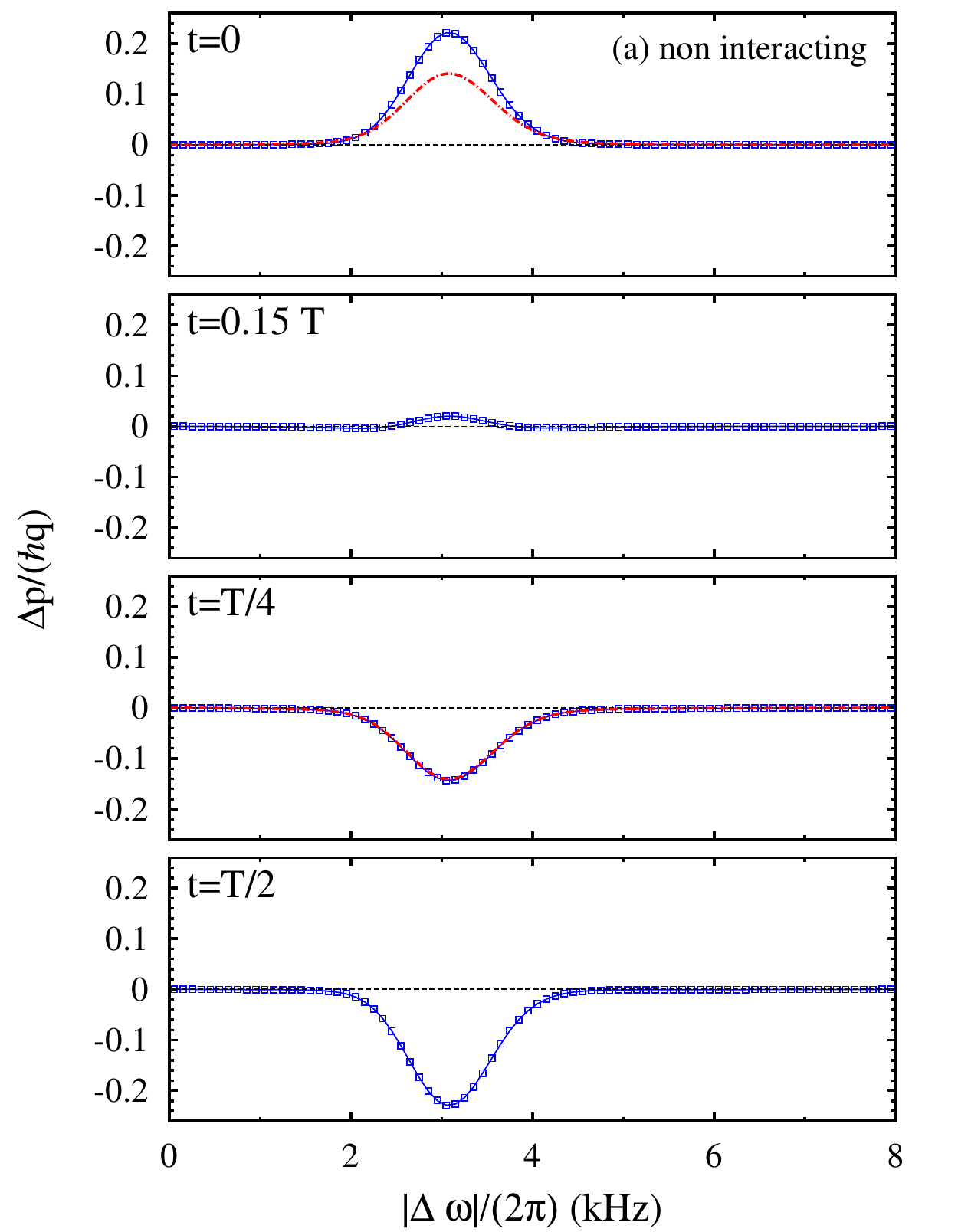}
\includegraphics[width= 0.49\columnwidth]{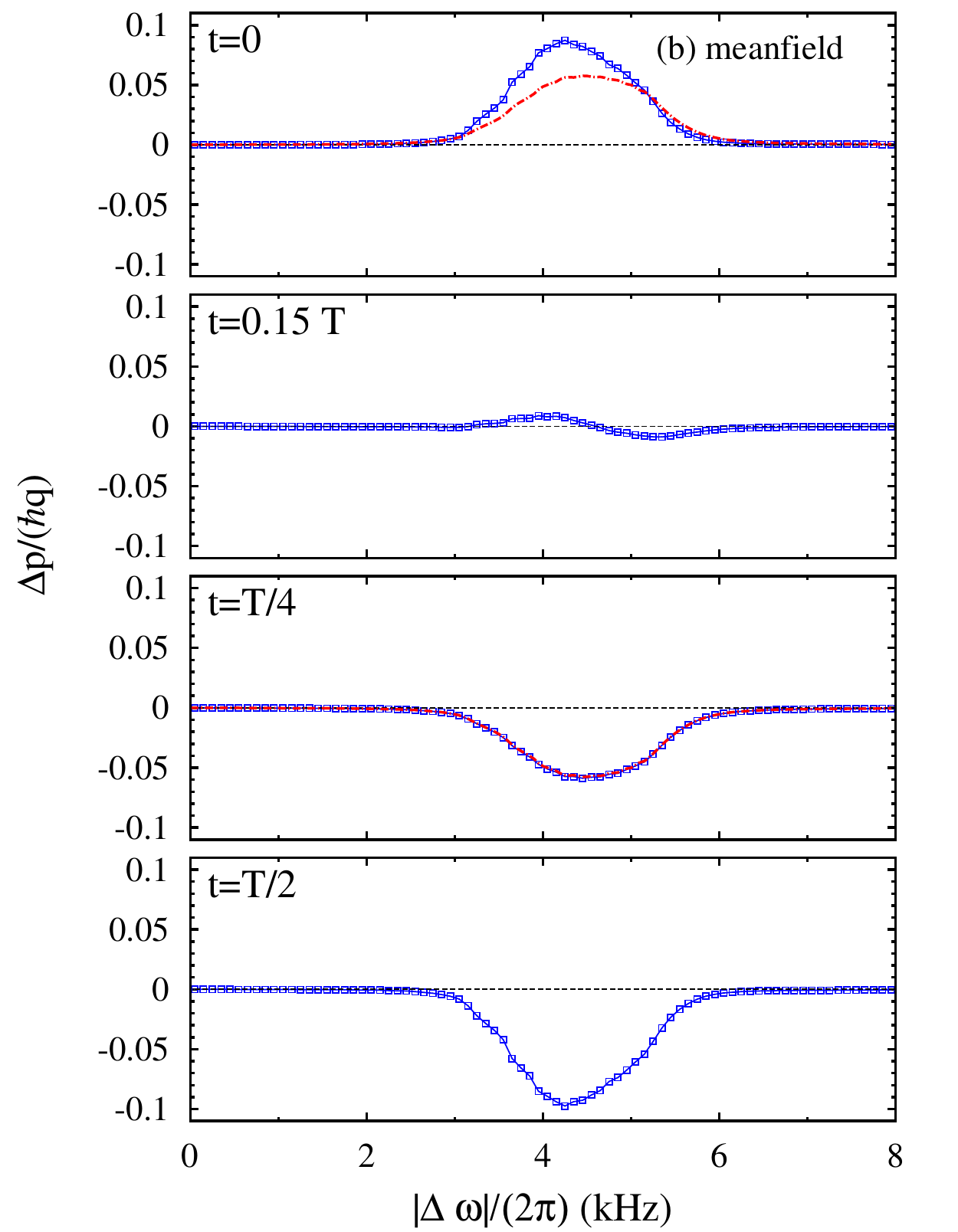}
\caption{{\bf In-trap dynamics of a condensate in the mean field regime.} Theoretical prediction for the value of $\Delta P/(\hbar q)$ (blue squares) as a function of the Bragg frequency for an array of quasi 1D BECs, for different holding times in the trap after the Bragg excitation: $t=0,0.15,0.25,0.5 T$ ($T$ being the trap period). The dotted-dashed (red) lines in the top panel represent the initial distributions of the center-of-mass $\langle x\rangle_{0}(|\Delta\omega|)$. The same line is shown inverted at $t=T/4$ (see text). The left and right columns correspond to (a) the non interacting and (b) interacting case in the mean-field regime, respectively.
}
\label{fig:gpe}
\end{figure}

As an example, let us consider the simple case of a Bose-Einstein condensate in a \textit{single}, quasi one-dimensional tube, in the mean-field regime. In this case, the response of the system to the Bragg pulse can be easily obtained by solving the following 1D Gross-Pitaevskii equation ($t\ge0$)
\begin{equation}
i\hbar\partial_{t}\psi=\left[-\frac{1}{2m}\nabla^{2}_{x} + V_{ho}(x) +\theta(t-t_B)V_{B}\cos(qx-\omega t) +g_{1D}|\psi|^{2}\right]\psi,
\end{equation}
where $g_{1D}=g/(2\pi a_{\perp})$, $g=4\pi\hbar^{2}a/m$ being the 3D interaction strength, $a$ the  scattering length for $^{87}$Rb, and $a_{\perp}=\sqrt{\hbar/(m\omega_{\perp})}$ the oscillator length in the transverse directions. In this specific example we consider $V_{B}/h=120$ Hz, $t_{B}=3$ ms, $\omega_{x}=2\pi\times60$ Hz, $\omega_{\perp}=2\pi\times42$ kHz, and an array of tubes that correspond\new{s} to the typical experimental configuration. The response of the system at different evolution times in the trap is shown in Fig. \ref{fig:gpe}, where the meanfield predictions are also compared to the non-interacting case \cite{nota3}. This figure shows that indeed, as follows from Eq. (\ref{ho}), the response patter at $t=T/2$ is reversed with respect to that at $t=0$, the evolution being periodic in time. For intermediate times, the shape of the signal is non trivial, depending on the relative weight and on the specific shape of the transferred momentum and the center of mass position as a function of the Bragg frequency $|\Delta\omega|$, at $t=0$. In the non interacting case, $\langle p\rangle_{0}(|\Delta\omega|)$ and $\langle x\rangle_{0}(|\Delta\omega|)$ are centered at the same value and almost symmetric around that point, so that the same symmetry property is preserved during the evolution. Instead, the response of an interacting condensate \new{is} characterized by a distribution of the center-of-mass position that is peaked at higher frequency with respect to the corresponding transferred momentum, and this affects dramatically the shape at intermediate times. In particular, at $t\simeq 0.15 T$ the signal shape has a characteristic sinusoidal-like form, whereas at $t=0.25 T$ it corresponds exactly to the reverse of the initial center-of-mass distribution $\langle x\rangle_{0}(|\Delta\omega|)$. Note that the signal at $t\simeq 0.15 T$ is analogous to that observed experimentally at $t=4$ ms, see Fig. \ref{spettridiversitempi}, though in that case the signal first reverses in the low frequency region. We attribute this difference to the fact that the experiment is performed in the strong-coupling regime, so that the response of the system to the Bragg perturbation is expected to be substantially different (we remark that a precise simulation of the dynamics of strongly correlated 1D systems under the effect of a Bragg perturbation can be very demanding, see e.g. \cite{cazalillaRMP2011}).

\section{Conclusions}
\label{sec:conclusions}

In conclusion, we have investigated the response of an array of 1D gases, comparing energy and momentum transfer in Bragg spectroscopy experiments. In the presence of an external trapping potential along the axis of the tubes, even if increasing the pulse time enhances the spectral resolution, the presence of the trap in principle provides an upper limit to the pulse duration. In addition, our experiment reveals that, in a regime of parameters well described by the linear response theory, and for time-duration of the Bragg perturbation smaller than a quarter of the trapping period, the proportionality relation between the momentum transfer and the dynamical structure factor is well respected. Moreover, we show that the in-trap dynamics during the Bragg pulse affects noticeably the response of the system. This analysis can be useful for interpreting the results of scattering experiments also in other more complex settings of ultracold gases in optical lattices or disordered potentials.

\section*{Acknowledgments}
We would like to acknowledge L. Fallani for critical reading of the manuscript. This work has been supported by ERC through the Advanced Grant DISQUA (Grant 247371), and by EC through EU FP7 the Integrated Project SIQS (Grant 600645), Universidad del Pais Vasco/Euskal Herriko Unibertsitatea under Programs No. UFI 11/55, Ministerio de Econom\'ia y Competitividad through Grant No. FIS2012-36673-C03-03, and the Basque Government through Grant No. IT-472-10.


\begin{thebibliography}{30}
\bibitem{nota1} Here, we have assumed $\sigma_x=\sigma_z$ owing to the cylindrical symmetry of the system.
\bibitem{nota3} The value of the amplitude $V_{B}/h$ of the Bragg pulse in the simulations is chosen in order to lie in the linear response regime for the non-interacting case with a pulse length of $t_B=3$ ms.
\end{thebibliography}
\end{document}